# The new Excellence Indicator in the World Report of

# the SCImago Institutions Rankings 2011

Lutz Bornmann $, Felix de Moya Anegón §, Loet Leydesdorff*

$ Max Planck Society, Administrative Headquarters, Hofgartenstr. 8, 80539

Munich, Germany.

§ CSIC/CCHS/IPP, Albasanz 26 Madrid, Spain.

* Amsterdam School of Communication Research, University of

Amsterdam, Kloveniersburgwal 48, NL-1012 CX, Amsterdam, The

Netherlands.

The SCImago Institutions Rankings (SIR) World Reports are published annually. These reports contain an international ranking of more than 2000 research institutions and organizations. The indicator values are based on publication and citation data from Scopus (Elsevier) for research-devoted institutions with at least 100 papers published within the year under study.

The second edition published in 2010 included four indicators for each institution: (1) publication output, (2) the percentage of output produced in collaboration with foreign institutions (international collaboration), (3) the ratio between the average scientific impact of an institution and the world average impact of papers published in the same time period and subject area (normalized impact), (4) the ratio of papers which an institution publishes in the most influential scholarly journals of the world (high-quality publications).

Subject areas are defined by Elsevier in Scopus on the base of journal classifications. The delineations in terms of subject areas provided by Scopus are continuously improved by SCImago using, among other things, the categorizations of the Web-of-Science and Medline (Lopez-Illescas, de Moya-Anegon, & Moed, 2008; López-Illescas, Noyons, Visser, De Moya-Anegón, & Moed, 2009). However, journal classifications remain a reduction of the complexity of inter-journal and inter-discipline relations which cannot be expected to match one-by-one (Boyack & Klavans, 2011; Rafols & Leydesdorff, 2009). The use of index terms (e.g., MeSH terms) provided by discipline-specific data bases (e.g., MEDLINE, US National Library of Medicine) where papers are classified to subject areas on a paper-by-paper basis might be more appropriate than journal classifications (Bornmann, Mutz, Neuhaus, & Daniel, 2008; Leydesdorff, 2006), but these classifications have not been available for large-scaled discipline-overlapping analyses hitherto.

Recently, the third edition of the World Report (available at http://www.scimagoir.com/pdf/sir_2011_world_report.pdf) was published with a new "Excellence Indicator" added. This indicator can be traced back to the methodological developments of Bornmann and Leydesdorff (2011) and Leydesdorff, Bornmann, Mutz, and Opthof (2011). The Excellence Indicator provides the percentage of papers published by an institution belonging to the top-10% papers in terms of numbers of citations, normalized for the same field of publications and the same publication year. Tijssen, Visser, and van Leeuwen (2002) and Tijssen and van Leeuwen (2006) argued that the top-10% of papers with the highest citation counts in a publication set can be considered as highly cited (see also Lewison, Thornicroft, Szmukler, & Tansella, 2007).

For example, an Excellence Indicator of 22% for an institution means that 22% of its papers belong to the top-10% most-highly-cited papers among those published in the same year and subject area (e.g., Biochemistry, Genetics & Molecular Biology; Immunology & Microbiology). SCImago uses an inclusive definition of the top-10%: when a set of documents has the same number of citations as the last document of the 10% core, these documents are all considered as part of the top-10% set. In some cases, the top-10% set is thus larger than 10%, but this is usually within the rounding of the first decimal.

The indicator can be considered as an item-oriented field-normalized citation score because each paper in an institutional publication set is analyzed whether it belongs to the top-10% of papers in the set of papers (covered by Scopus) with the same publication year and subject area. However, different from normalizations based on average values (Lundberg, 2007; Opthof & Leydesdorff, 2010; Waltman, van Eck, van Leeuwen, Visser, & van Raan, 2011), the top-10% can be considered as a non-parametric statistics. This non-parametric approach accounts for the prevailing skewedness of citation distributions (Albarrán, Crespo, Ortuño, & Ruiz-Castillo, 2011; Seglen, 1992).

The Excellence Indicator has two advantages: First, the percentage for an institution (the observed number) can be compared with the reference value (expected value) of 10%. The expected number in the top-10% for a set of papers selected at random would be 10% (Agarwal

& Searls, 2009; Bornmann & Mutz, 2011). Accordingly, institutions in the World Report with percentages above 10% perform above expectation (or, in other words, above the reference standard), and institutions with percentages below 10% perform below expectation. The percentages of different institutions (and their deviations from 10%) can be compared directly with one another since these ratios were already normalized for respective publication years and subject areas.

Secondly, the Excellence Indicator allows for testing whether (1) the difference between the institution's percentage and the expected value of 10% or (2) the percentage difference between two institutions are statistically significant. The statistical significance test analyzes whether the difference (e.g., between the observed and expected institution's number of top-10% papers) which is reached on the base of a sample (e.g., papers published between 2003 and 2007) is valid (in all likelihood) for all (ever published) papers of the institute in question (covered by Scopus) (Bornmann *et al*., 2008). If the test is statistically significant the difference does not seem to be a random event but can be interpreted beyond the analyzed sample data.

The appropriate test is the *z*-test for two independent proportions (Sheskin, 2007, pp. 637-643). This test can be used for evaluating both the degree to which an observed number differs from the expected number and whether the observed numbers for two institutions differ, respectively (Bornmann & Leydesdorff, 2011). In general, the test statistics can be formulated as follows:

$$z = \frac{p_1 - p_2}{\sqrt{p(1-p)\left[\frac{1}{n_1} + \frac{1}{n_2}\right]}}$$

where: $n_1$ and $n_2$ are the numbers of all papers published by institutions 1 and 2 (under the column "Output" in the World Report); and $p_1$ and $p_2$ are the values of the Excellence Indicators of institutions 1 and 2. Furthermore:

$$p = \frac{t_1 + t_2}{n_1 + n_2}$$

where: $t_1$ and $t_2$ = the numbers of top-10% papers of institutions 1 and 2. These numbers can be calculated on the base of "Output" and "Excellence Indicator". In the case of testing observed versus expected for the same set, $n_1 = n_2$; $p_1$ is the observed value of the Excellence Indicator and $p_2$ the expected value which is for stochastic reasons: 10% of $n_2$.

An absolute value of $z$ larger than 1.96 indicates statistical significance of the difference between the two proportions at the five percent level ($p<.05$); the critical value for a test at the one-percent level ($p<.01$) is 2.576. If a reader of the World Report conducts a series of tests for many institutions, a higher significance level than five percent may have to be chosen. There is a possibility of family-wise accumulation of Type-I errors (Leydesdorff *et al.*, 2011).

For example, at the 17th position in the SCImago Institutions Rankings, University of California, Los Angeles (UCLA) has an output of 37,994 papers with an excellence rate of 28.9%. Stanford University follows at the 19th position with 37,885 papers and a 29.1% excellence rate. Using the above formulas, $z = - 0.607$. The difference between these two institutions thus is *not* statistically significant. A calculator is provided at http://www.leydesdorff.net/scimago11/scimago11.xls in which one can fill out this test for the comparison of any two institutions and also for each institution on whether it scores significantly above or below expectation (assuming that 10% of the papers are for stochastic reasons in the top-10% set).

As the interpretations and calculations described in this Letter to the Editor show, the simple percentage of top-10% papers for an institution – the new Excellence Indicator – offers already a lot of possibilities for the comparison of an institution against an expectation or reference standard, and with other institutions by using non-parametric statistics for testing significance. Leydesdorff and Bornmann (2011) further developed this statistics to the *Integrated Impact Indicator* (*I3*) which allows for more refinement of the choices, but this measure is perhaps less intuitively easy to understand than the top-10% for a non-specialist audience (Bornmann & Leydesdorff, in press).